\documentclass[11pt]{article}
\usepackage{graphicx,caption,subfigure}
\makeatletter
\newcommand{\singlespacing}{\let\CS=\@currsize\renewcommand{\baselinestretch}{1.0}\tiny\CS}
\newcommand{\doublespacing}{\let\CS=\@currsize\renewcommand{\baselinestretch}{1.5}\tiny\CS}

\begin{document}
\title{Light Hadron Production in Proton-Proton Collisions at Different LHC Energies: Measured Data versus a Model\thanks{The work is supported by the University Grants Commission of India under Grant No PSW-30/12(ERO) dt.05 Feb-13.}}
\author{P.
Guptaroy$^1$\thanks{e-mail: gpradeepta@rediffmail.com(Corresponding Author) }, S. Guptaroy$^2$\thanks{e-mail: simaguptaroy@yahoo.com}\\
{\small  $^1$Department of Physics, Raghunathpur College,}\\
 {\small P.O.: Raghunathpur 723133,  Dist.: Purulia (WB), India.}\\
 {\small $^2$ Department of Physics, Basantidevi College,}\\
 {\small 147B Rashbehari Avenue, Kolkata 700029 India.}
 }
\date{}
\maketitle
\bigskip
\bigskip
\begin{abstract}
Experiments involving proton-proton collisions at energies $\sqrt{s_{NN}}$ = 0.9, 2.76 and 7 TeV in Large Hadron Collider (LHC) have produced a vast amount of high-precision data. Here, in this work, we have chosen to analyse the two aspects of the measured data, viz., (i) the $p_T$ -spectra of pions, kaons, proton-antiproton at above-mentioned energies, and (ii) some of their very important ratio-behaviours, in the light of a version of the Sequential Chain Model (SCM). The agreements between the measured data and model-based results are generally found to be modestly satisfactory.
\end{abstract}
\bigskip
 {\bf{Keywords}}: Relativistic heavy ion collisions, baryon production, light mesons
\par
 {\bf{PACS nos.}}: 25.75.-q, 13.60.Rj, 14.40.Be
\newpage
The production of hadronic particles in high energy heavy ion physics is of special
interest to understand the underlying mechanisms leading to such reaction products and to test
the predictions from non-perturbative QCD processes. The yield of identified hadrons, their
multiplicity distributions, the rapidity and transverse momentum spectra are the
basic observables in heavy ion collisions. The measurement of such global observables has produced
over the years, new and interesting insights about the involved production mechanisms, allowing in turn to
improve the theoretical description of multiparticle production processes in heavy ion collisions.
\par
In $pp$ collisions at ultra-relativistic energies $\sqrt{s_{NN}}$= 0.9, 2.76 and 7 TeV, the bulk of the particles produced at mid-rapidity
have transverse momenta below 2 GeV/c. First principles calculations based on perturbative
QCD are not able to provide detailed predictions of particle production.\cite{riggi} Here, our basic objective in the present work is to interpret a part of significant data, like the transverse momenta spectra and some ratio-behaviours of pions, kaons and protons, obtained from $pp$ collisions at LHC energies $\sqrt{s_{NN}}$= 0.9, 2.76 and 7 TeV, with the help of the Sequential Chain Model (SCM). The another goal of the present work is to put this alternative approach to a segment of LHC-data with a view to assessing the success(es)/failure(s) of it.
\par
According to
this Sequential Chain Model (SCM), high energy hadronic
interactions boil down, essentially, to the pion-pion interactions;
as the protons are conceived in this model as
$p$~=~($\pi^+$$\pi^0$$\vartheta$) where $\vartheta$ is a spectator particle needed for
the dynamical generation of quantum numbers of the nucleons \cite{pgr10}-\cite{bhat882}. The
production of pions in the present scheme occurs as follows: the
incident energetic $\pi$-mesons in the structure of the projectile
proton(nucleon) emits a rho($\varrho$)-meson in the interacting
field of the pion lying in the structure of the target proton, the
$\varrho$-meson then emits a $\pi$-meson and is changed into an
omega($\omega$)-meson, the $\omega$-meson then again emits a
$\pi$-meson and is transformed once again into a $\varrho$-meson and
thus the process of production of pion-secondaries continue in the
sequential chain of $\varrho$-$\omega$-$\pi$ mesons. The production mechanism is shown schematically in Fig. 1. The two ends of
the diagram contain the baryons exclusively \cite{pgr10}-\cite{bhat882}.
\par
In a similar fashion, the $K^-$($K^+$) and the baryon-antibaryon have been produced in the SCM and they are shown in Fig.2 and Fig.3 respectively.
 \par
 The fundamental expressions for final (analytical) calculations are
derived here on the basis of field-theoretic considerations and the
use of Feynman diagram techniques with the infinite momentum
frame tools and under impulse approximation method. The expressions for inclusive
cross-sections of the $\pi^-$, $K^-$ and $\bar p$-secondaries would pick up from \cite{pgr10}-\cite{bhat882} and they are given by the following relations;
\begin{equation}\displaystyle E\frac{d^3\sigma}{dp^3}|_{pp \rightarrow
\pi^- x}  \cong \Gamma_{\pi^-} \exp(- 2.38 <n_{\pi^-}>_{pp}
x)\frac{1}{p_T^{(N_R^{\pi^-})}} \exp(\frac{-2.68
p_T^2}{<n_{\pi^-}>_{pp}(1-x)})   ~ ,
\end{equation}
with \begin{equation}\displaystyle {<n_{\pi^+}>_{pp} ~ \cong  ~
<n_{\pi^-}>_{pp} ~ \cong
 ~ <n_{\pi^0}>_{pp}  ~ \cong  ~ 1.1s^{1/5} ~,}
 \end{equation}

\begin{equation}
\displaystyle E \frac{d^3\sigma}{dp^3}|_{pp \rightarrow K^- x}
 ~ \cong  ~ \Gamma_{K^-}\exp(  -  6.55 <n_{K^-}>_{pp}
 x) ~ \frac{1}{p_T^{(N_R^{K^-})}}\exp(\frac{-  1.33
  p_T^2}{<n_{K^-}>^{3/2}_{pp}}) ~ ~ ,
\end{equation}
 and with
\begin{equation}
\displaystyle <n_{K^+}>_{pp}   \cong  <n_{K^-}>_{pp}  \cong
  <n_{K^0}>_{pp}  \cong  <n_{\bar{K^0}}>_{pp} \cong
5\times10^{-2}  s^{1/4}  ,
\end{equation}

\begin{equation}
\displaystyle E\frac{d^3\sigma}{dp^3}|_{pp\rightarrow{\bar p}x}
 ~ \cong \Gamma_{\bar p} \exp(-25.4 <n_{\bar{p}}>_{pp} x)\frac{1}{p_T^{({N_R}^{\bar p})}}\exp(\frac{-0.66 ((p_T^2)_{\bar
p}+ {m_{\bar p}}^2)}{<n_{\bar p}>^{3/2}_{pp} (1-x)})
 ~  ,
\end{equation}
with $m_{\bar p}$
is the mass of the antiprotons. For ultrahigh energies
\begin{equation}
\displaystyle{ <n_{\bar p}>_{pp}  ~ \cong <n_p>_{pp}  ~ \cong
 ~ 2\times10^{-2} ~ s^{1/4} ~ .}
 \end{equation}

 where $\Gamma_{C^-}$ ($C^-$ stands for $\pi^-$, $K^-$ or $\bar p$)  is the
normalisation factor which will increase as the inelastic
cross-section increases and it is different for different energy
region and for various collisions.  The terms
$p_T$, $x$ in equations (1), (3) and (5) represent the transverse momentum,
Feynman Scaling variable respectively. Moreover, by definition, $x ~
= ~ 2p_L/{\sqrt s}$, where $p_L$ is the longitudinal momentum of the
particle. The $s$ in equations (2), (4) and (6)  is the square of the c.m. energy.
\par
$1/p_T^{N_R^{C^-}}$ represnts the `constituent
rearrangement term' arising out of the partons inside the proton
which essentially provides a damping term in terms of a power-law in
$p_T$ with an exponent of varying values depending on both the
collision process and the specific $p_T$-range. The choice of
${N_R}$ would depend on the following factors: (i) the specificities
of the interacting projectile and target, (ii) the particularities
of the secondaries emitted from a specific hadronic or nuclear
interaction and (iii) the magnitudes of the momentum transfers and
of a phase factor (with a maximum value of unity) in the
rearrangement process in any collision. And this is a factor for
which we shall have to parameterize alongwith some physics-based
points indicated earlier. The parametrization is to be done for two
physical points, viz., the amount of momentum transfer and the
contributions from a phase factor arising out of the rearrangement
of the constituent partons. Collecting and combining all these,
the relation is to be given by \cite{pgr08}
\begin{equation}\displaystyle
N_R=4<N_{part}>^{1/3}\theta,
\end{equation}
where $<N_{part}>$ denotes the average number of participating
nucleons and $\theta$ values are to be obtained phenomenologically
from the fits to the data-points.
\par
 On the average, the particles are produced in
charge-independent equal measure, for which roughly one-third of the
particles could be reckoned to be positively charged, one third are
negatively charged and the rest one third are neutral. But,
according to the present mechanism of particle production, there are
some specifically exclusive means to produce positive particles, of
which $\pi^+$, $K^+$, and $p$ are the members. They are produced
from within the structure of protons (nucleons). These production
characteristics and the quantitative expressions for their special
production have been dwelt upon in detail in Ref. \cite{bhat882}.
Let us assort the relevant expressions therefrom as results to be
used here.
\par
 For production of positive pions
[$\pi^+$ mesons] the excess term could be laid down
by the following expressions \cite{bhat882}:

\begin{equation}\displaystyle
(B_{\pi^+})_{pp} = {\frac {4}{3}}{g^2_{p \pi
\pi}}{\frac{(P'+K)^2}{[(P'+K)^2-m_p^2]^2}} A(\nu,
q^2)_{\pi}\int{\frac{d^3k_{\pi}}{2k_0(2 \pi)^3}\exp(-ik_{\pi}x)},
\end{equation}
where the symbols have their contextual connotation with the
following hints to the physical reality of extraneous $\pi^+$, as
non-leading secondaries. The first parts of the above equations
(Eqn.(8)), contain the coupling strength parameters, the second
terms of the above equations are just the propagator for excited
nucleons. The third terms represent the common multiparticle
production amplitudes along with extraneous production modes and the
last terms indicate simply the phase space integration terms on the
probability of generation of a single $\pi^+$. These expressions are
to be calculated by the typical field-theoretical techniques and are
to be expressed -- if and when necessary -- in terms of the relevant
variable and/or measured observables.
\par
In order to arrive at the transverse momentum distribution of
$\pi^+$, one has to consider the Eqn. (1), along with eqn. (8). For
excess $\pi^+$ production, a factor represented by $(1+\gamma^{\pi^+}
p_T^{\pi^+})$ is to be operated on inclusive cross-section as an
multiplier \cite{bhat882}.
 \par
Adopting the above procedure, as we indicated for the production of
positive pions, we obtain for the transverse momentum distribution
of $K^+$ a multiplicative factor $\sim (1+\gamma^{K^+} p_T^{K^+})$ to be
operated on the inclusive cross-section of $K^-$ production.
\cite{bhat882}.
\par
Similarly, for the production of protons, we
obtain for the transverse momentum distribution of $p$ by operating
a multiplicative factor $\sim (1+\gamma^p p_T^{p})$, on inclusive cross-section of $\bar p$.
\par
We will dwell upon here properties of the nature of $p_T$-spectra and some particle-production
ratios of some light hadrons, like $\pi^\pm$, $K^\pm$, $\bar p$ and $p$, in $pp$-collisions at energies $\sqrt{s_{NN}}$= 0.9, 2.76 and 7 TeV in LHC.
\par
The general form of our SCM-based transverse-momentum distributions
for $p+p\rightarrow C^-+X$-type reactions can be written in the
following notation:
\begin{equation}\displaystyle{
\frac{1}{2\pi p_T}  \frac{d^2N_{C}}{d\eta dp_T}|_{p+p\rightarrow
C^-+X}=\alpha_{C^-}\frac{1}{p_T^{N_R^{C^-}}}\exp(-\beta_{C^-} \times
p_T^2).}
\end{equation}
The values of $\alpha_{\pi^-}$ and $\beta_{\pi^-}$ , for example, can be calculated from the following relations [equation 1]:
\begin{equation}\displaystyle{
\alpha_{\pi^-}=\Gamma_{\pi^-}\exp(- 2.38
<n_{\pi^-}>_{pp} x)}
\end{equation}
\begin{equation}\displaystyle{
\beta_{\pi^-}=\exp(\frac{-2.68
}{<n_{\pi^-}>_{pp}(1-x)})}
\end{equation}
And for calculation of $(N_R^{\pi^-})_{pp}$, we use eqn.(7). The values of $(\alpha_{C^-})_{pp}$, $(N_R^{C^-})_{pp}$ and $(\beta_{C^-})_{pp}$ for $K^-$ and $\bar p$ can be calculated in a similar way by using equations (3)-(6).
\par
The values of $(\alpha_{C^-})_{pp}$, $(N_R^{C^-})_{pp}$ and $(\beta_{C^-})_{pp}$ ($C^-$ stands for $\pi^-$, $K^-$ or $\bar p$ respectively.)  for different energies are given in the left panels of Table 1. The experimental data for the inclusive cross-sections versus $p_T$ [GeV/c] for $\pi^-$, $K^-$ and $\bar p$ production in $p+p$
 interactions at $\sqrt{s_{NN}}$ = 0.9, 2.76 and 7 TeV are taken
from Ref. \cite{cms12} and they are plotted in Figs. 4(a), 4(c) and 4(e) respectively. The solid
lines in those figures depict the SCM-based plots while the dotted lines show the Pythia-induced calculations.
\par
In order to arrive at the transverse momentum distribution of
$\pi^+$, $K^+$ and $p$ one has to consider the Eqn. (1), eqn. (4), eqn. (6), eqn. (8), eqn. (9) along with eqn. (10). For
excess $\pi^+$ production, the factor  $(1+\gamma^{\pi^+}
p_T^{\pi^+})$ is to be operated on $\frac{1}{2\pi
p_T}\frac{d^2N}{dp_Tdy}|_{p+p\rightarrow\pi^-+ X}$ as an
multiplier. $\gamma^{\pi^+}\simeq (20\pi
g^2_{{\rho}{\pi}{\pi}}/<n_{\pi}>)/\sqrt{s}\simeq 0.44$ \cite{pgr07}. Taking $<p_T>_{\pi^+} \simeq 0.31$ GeV/c
\cite{phobos08}, the calculated values of $\alpha_{\pi^+}$
 for
different energies are given in the right panel of Table 1. The values of $(N_R^{\pi^+})_{pp}$ and $(\beta_{\pi^+})_{pp}$ are remain same and they are given in the right panel of Table 1.
\par
 Similarly, we obtain for the transverse momentum distribution of $K^+$, the multiplicative factor $\sim (1+\gamma^{K^+} p_T^{K^+})$ is to be operated on $\frac{1}{2\pi
p_T}\frac{d^2N}{dp_Tdy}|_{p+p\rightarrow K^-+ X}$ as an
multiplier.The $\gamma^{K^+}$ has been calculated
 $\gamma^{K^+}\simeq (4\pi g^2_{K N \Lambda} + 4\pi g^2_{\Sigma K N})/2\sqrt{s}\simeq 0.082$
\cite{pgr07}. We use the value of $<p_T>_{K^+} \simeq
0.36$ GeV/c \cite{phobos08}. The right panel of Table 1 depicts all calculated values of $\alpha_{K^+}$,
$N_R^{K^+}$ and $\beta_{K^+}$.
\par
And we
obtain for the transverse momentum distribution of $p$ by operating
a multiplicative factor $\sim (1+\gamma^p p_T^{p})$ on $\frac{1}{2\pi p_T}\frac{d^2N}{dp_Tdy}$.
The value of $\gamma^p\sim 0.32$ \cite{pgr07} and by taking $<p_T>_p
\simeq 0.50$ GeV/c \cite{phobos08}. The right panel of Table 1 depicts all calculated values of $\alpha_{p}$,
$N_R^{p}$ and $\beta_{p}$ for energies $\sqrt{s_{NN}}$ = 0.9, 2.76 and 7 TeV. In Figures 4(b), 4(d) and
4(f), we have plotted experimental versus theoretical results for
$\pi^+$ , $K^+$ and $p$ production in $p+p$ collisions at energies $\sqrt{s_{NN}}$ = 0.9, 2.76 and 7 TeV, respectively. Data are taken from the
Refs. \cite{cms12}. The solid lines in those
Figures are the SCM-based plots while the dotted lines show the PYTHIA-induced calculations.
\par
The $(K^++K^-)/(\pi^++\pi^-)$ and $(p+\bar p)/(\pi^++\pi^-)$ ratios at energies $\sqrt{s_{NN}}$ = 0.9, 2.76 and 7 TeV have been calculated from eqn. (9) and Table 1 and the results are plotted in Figs. 5(a), 5(c) and 5(e). Data are taken from the
Refs. \cite{cms12} and \cite{preg}. Similarly, $\pi^-/\pi^+$, $K^-/K^+$ and $\bar p/p$-ratios are plotted in the right panel of Fig. (5), i.e., Figs. 5(b), 5(d) and 5(f). Data are taken from the Refs. \cite{cms12}. The solid lines in those Figures are the average SCM-based plots.
\par
Let us make some general observations and specific comments on a case-to-case basis.
\par
1) The very basic model used here is essentially of non-standard
type. The measures of invariant yields against transverse momenta ($p_T$) obtained on the basis of the SCM for pions, kaons and protons at LHC energies $\sqrt{s_{NN}}$= 0.9, 2.76 and 7 TeV  are depicted in Fig. (4). The calculated values of $(\alpha_{C})_{pp}$, $(N_R^{C})_{pp}$ and $(\beta_{C})_{pp}$ ($C$ stands for $\pi$, $K$ or $p$ respectively.) of eqn.(9) have been shown in Table 1. Besides, we compare the model-based calculations with the PYTHIA-based results. Comparisons show neither sharp disagreement nor any good agreement between these two. Results show a modest degree of success.
\par
There are some disagreements of the model in describing the data in the low-$p_T$ region. These are due to the fact that the model has turned essentially into a mixed one with the inclusion of power law due to the inclusion of partonic rearrangement factor. This power law term disturbs, to a considerable extent, the agreement between the data and the model. However, the power-law part of the equation might not be the only factor for this type of discrepancy. The initial condition and dynamical evolution in heavy-ion collisions are more complicated than we expect. Till now, we do not know the exact nature of reaction mechanism. One might take into account some other factors like radial flow or thermal equilibrium. The thermal equilibrium was included in blast-wave model in a recent paper. \cite{ma} But we are not able to include these factors. The model is, thus, different from thermal model and blast-wave model \cite{ma} in some way. These are the factors that the model is lagging.
\par
2) The calculated values of $K/\pi$ and $p/\pi$ are compared with the experimental ones and they are plotted in Figs. 5(a), 5(c) and 5(e). The $K/\pi$ ratios show a modest degree of success while for $p/\pi$, the theoretical values differ slightly from the experimental ones in the low-$p_T$ part. In a similar way, $\pi^-/\pi^+$, $K^-/K^+$ and $\bar p/p$-ratios are plotted in the right panel of Fig. (5). Here, the model reproduce the data modestly well.
\par
One point needs to be addressed here. The $K^-/K^+$-ratios at $\sqrt{s_{NN}}$= 0.9, 2.76 and 7 TeV are calculated from Table 1 and they are $\sim$ 0.98, 0.98 and 0.98 respectively. This is due to extraneous mode of production of positive particles. No hyperon has been produced in the present scheme of the SCM. The production mechanism of kaons are shown schematically in Fig. 2.
\par
 However, on the overall basis, the total approach might provide an alternative route to understand and interpret the
behaviour of high energy collisions.   \\
{\bf{Acknowledgements}}\\
The authors are very much thankful to the learned Referees for their valued comments and constructive criticisms.

\newpage
\begin{table}
\singlespacing \caption{Values of $\alpha$, $N_R$ and $\beta$ for
pions, kaons, antiproton and proton productions in $p+p$ collisions at
$\sqrt{s_{NN}}$=0.9, 2.76 and 7 TeV}
\begin{center}
\begin{tabular}{c}
\hline \begin{tabular}{c|c} $\pi^-$& $\pi^+$\\
\end{tabular}\\
$\sqrt{s_{NN}}$=0.9 TeV\\
 \begin{tabular}{c|c} \begin{tabular}{ccc}
$\alpha_{\pi^-}$&$N_R^{\pi^-}$&$\beta_{\pi^-}$\\
\end{tabular}
&\begin{tabular}{ccc}
$\alpha_{\pi^+}$&$N_R^{\pi^+}$&~$\beta_{\pi^+}$\\
\end{tabular}\\
\end{tabular}\\
\hline \begin{tabular}{c|c} \begin{tabular}{ccc}
0.310&2.054&0.128\\
\end{tabular}
&\begin{tabular}{ccc}
0.350&2.054&0.128\\
\end{tabular}\\
\end{tabular}\\
\hline
$\sqrt{s_{NN}}$=2.76 TeV\\
  \begin{tabular}{c|c} \begin{tabular}{ccc}
$\alpha_{\pi^-}$&$N_R^{\pi^-}$&$\beta_{\pi^-}$\\
\end{tabular}
&\begin{tabular}{ccc}
$\alpha_{\pi^+}$&$N_R^{\pi^+}$&$\beta_{\pi^+}$\\
\end{tabular}\\
\end{tabular}\\
\hline \begin{tabular}{c|c} \begin{tabular}{ccc}
0.461&2.163&0.085\\
\end{tabular}
&\begin{tabular}{ccc}
0.521&2.163&0.085\\
\end{tabular}\\
\end{tabular}\\
 \hline
$\sqrt{s_{NN}}$=7 TeV\\
  \begin{tabular}{c|c} \begin{tabular}{ccc}
$\alpha_{\pi^-}$&$N_R^{\pi^-}$&$\beta_{\pi^-}$\\
\end{tabular}
&\begin{tabular}{ccc}
$\alpha_{\pi^+}$&$N_R^{\pi^+}$&$\beta_{\pi^+}$\\
\end{tabular}\\
\end{tabular}\\
\hline \begin{tabular}{c|c} \begin{tabular}{ccc}
0.602&2.183&0.075\\
\end{tabular}
&\begin{tabular}{ccc}
0.683&2.183&0.075\\
\end{tabular}\\
\end{tabular}\\
\hline
\end{tabular}

\begin{tabular}{c}
\hline \begin{tabular}{c|c} $K^-$&$K^+$\\
\end{tabular}\\
$\sqrt{s_{NN}}$=0.9 TeV\\
 \begin{tabular}{c|c} \begin{tabular}{ccc}
$\alpha_{K^-}$&$N_R^{K^-}$&$\beta_{K^-}$\\
\end{tabular}
&\begin{tabular}{ccc}
$\alpha_{K^+}$&$N_R^{K^+}$&$\beta_{K^+}$\\
\end{tabular}\\
\end{tabular}\\
\hline \begin{tabular}{c|c} \begin{tabular}{ccc}
0.130&1.354&0.248\\
\end{tabular}
&\begin{tabular}{ccc}
0.133&1.354&0.248\\
\end{tabular}\\
\end{tabular}\\
\hline
$\sqrt{s_{NN}}$=2.76 TeV\\
  \begin{tabular}{c|c} \begin{tabular}{ccc}
$\alpha_{K^-}$&$N_R^{K^-}$&$\beta_{K^-}$\\
\end{tabular}
&\begin{tabular}{ccc}
$\alpha_{K^+}$&$N_R^{K^+}$&$\beta_{K^+}$\\
\end{tabular}\\
\end{tabular}\\
\hline \begin{tabular}{c|c} \begin{tabular}{ccc}
0.145&1.414&0.148\\
\end{tabular}
&\begin{tabular}{ccc}
0.147&1.414&0.148\\
\end{tabular}\\
\end{tabular}\\
 \hline
$\sqrt{s_{NN}}$=7 TeV\\
  \begin{tabular}{c|c} \begin{tabular}{ccc}
$\alpha_{K^-}$&$N_R^{K^-}$&$\beta_{K^-}$\\
\end{tabular}
&\begin{tabular}{ccc}
$\alpha_{K^+}$&$N_R^{K^+}$&$\beta_{K^+}$\\
\end{tabular}\\
\end{tabular}\\
\hline \begin{tabular}{c|c} \begin{tabular}{ccc}
0.195&1.454&0.134\\
\end{tabular}
&\begin{tabular}{ccc}
0.198&1.454&0.134\\
\end{tabular}\\
\end{tabular}\\
\hline
\end{tabular}

\begin{tabular}{c}
\hline \begin{tabular}{c|c} $\bar p$&$p$\\
\end{tabular}\\
$\sqrt{s_{NN}}$=0.9 TeV\\
 \begin{tabular}{c|c} \begin{tabular}{ccc}
$\alpha_{\bar p}$&$N_R^{\bar p}$&$\beta_{\bar p}$\\
\end{tabular}
&\begin{tabular}{ccc}
$\alpha_{p}$&$N_R^{p}$&$\beta_{p}$\\
\end{tabular}\\
\end{tabular}\\
\hline \begin{tabular}{c|c} \begin{tabular}{ccc}
0.073&1.154&0.248\
\end{tabular}
&\begin{tabular}{ccc}
0.085&1.154&0.248\\
\end{tabular}\\
\end{tabular}\\
\hline
$\sqrt{s_{NN}}$=2.76 TeV\\
  \begin{tabular}{c|c} \begin{tabular}{ccc}
$\alpha_{\bar p}$&$N_R^{\bar p}$&$\beta_{\bar p}$\\
\end{tabular}
&\begin{tabular}{ccc}
$\alpha_{p}$&$N_R^{p}$&$\beta_{p}$\\
\end{tabular}\\
\end{tabular}\\
\hline \begin{tabular}{c|c} \begin{tabular}{ccc}
0.095&1.243&0.148\\
\end{tabular}
&\begin{tabular}{ccc}
0.108&1.243&0.148\\
\end{tabular}\\
\end{tabular}\\
 \hline
$\sqrt{s_{NN}}$=7 TeV\\
  \begin{tabular}{c|c} \begin{tabular}{ccc}
$\alpha_{\bar p}$&$N_R^{\bar p}$&$\beta_{\bar p}$\\
\end{tabular}
&\begin{tabular}{ccc}
$\alpha_{p}$&$N_R^{p}$&$\beta_{p}$\\
\end{tabular}\\
\end{tabular}\\
\hline \begin{tabular}{c|c} \begin{tabular}{ccc}
0.120&1.263&0.134\\
\end{tabular}
&\begin{tabular}{ccc}
0.130&1.263&0.134\\
\end{tabular}\\
\end{tabular}\\
\hline
\end{tabular}
\end{center}
\end{table}
\newpage
\begin{figure}
\centering
\includegraphics[width=4.0in]{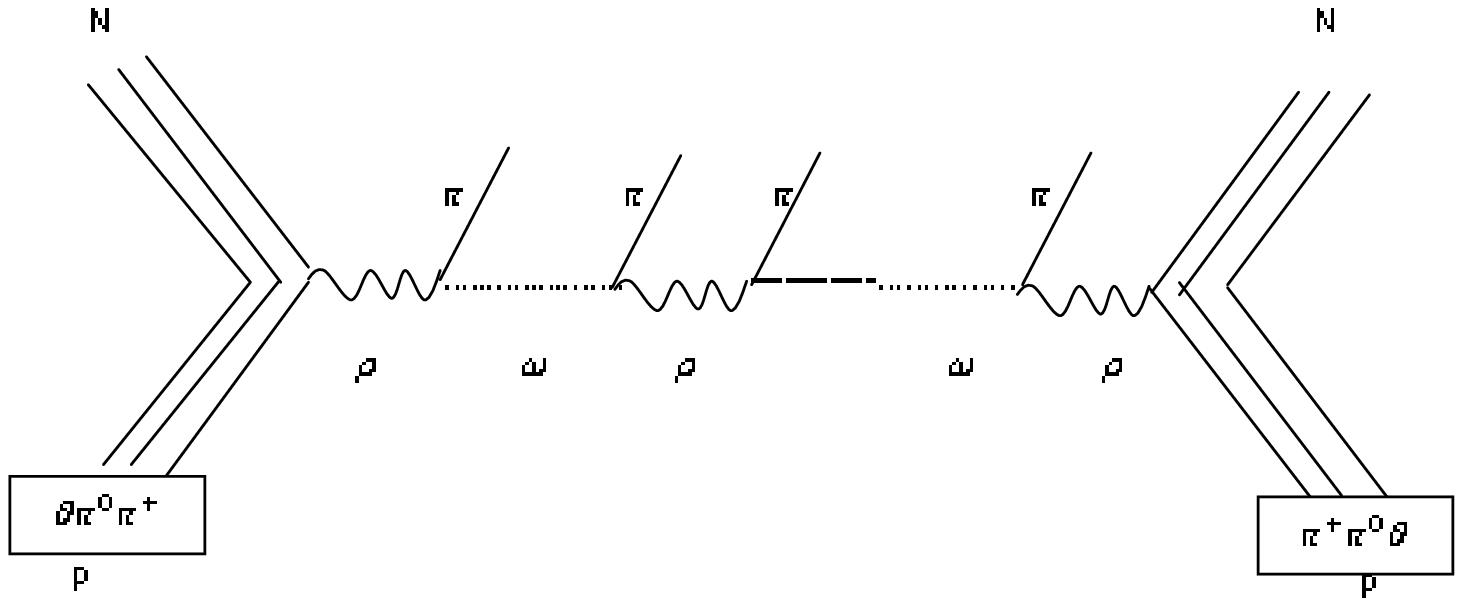}
\caption{Feynman diagrams for the production of $\pi$ in the Sequential Chain Model}
\end{figure}
 \begin{figure}
\centering
 \includegraphics[width=4.0in]{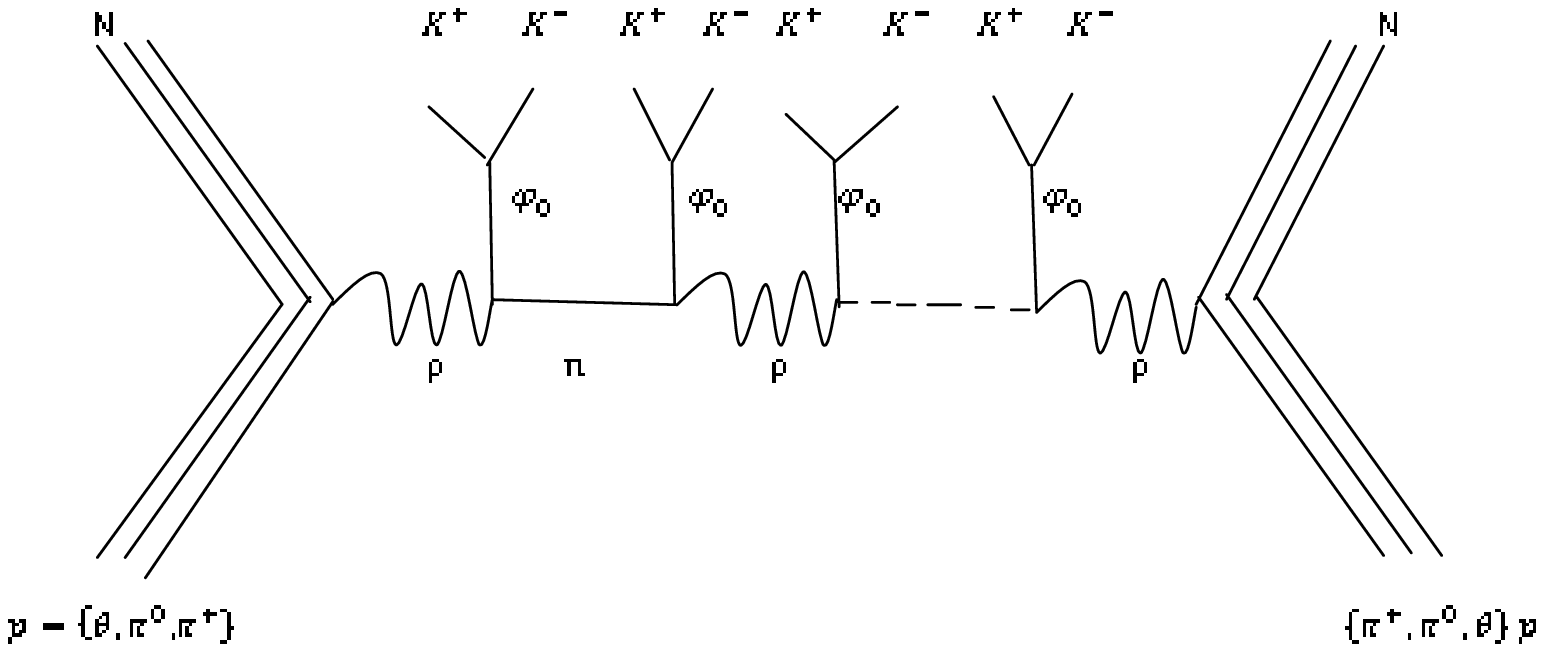}
  \caption{Feynman diagrams for the production of $K^\pm$ in the Sequential Chain Model}
\end{figure}
 \begin{figure}
\centering
 \includegraphics[width=4.0in]{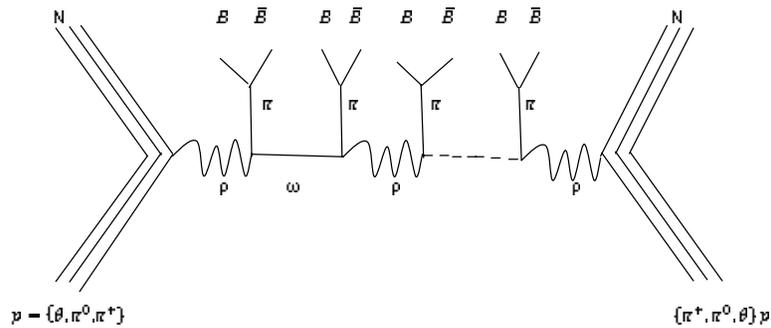}
\caption{\small  Feynman diagrams for the production of $\bar p$ and $p$ in the Sequential Chain Model.}
\end{figure}
\begin{figure}
\subfigure[]{
\begin{minipage}{.5\textwidth}
\centering
\includegraphics[width=2.5in]{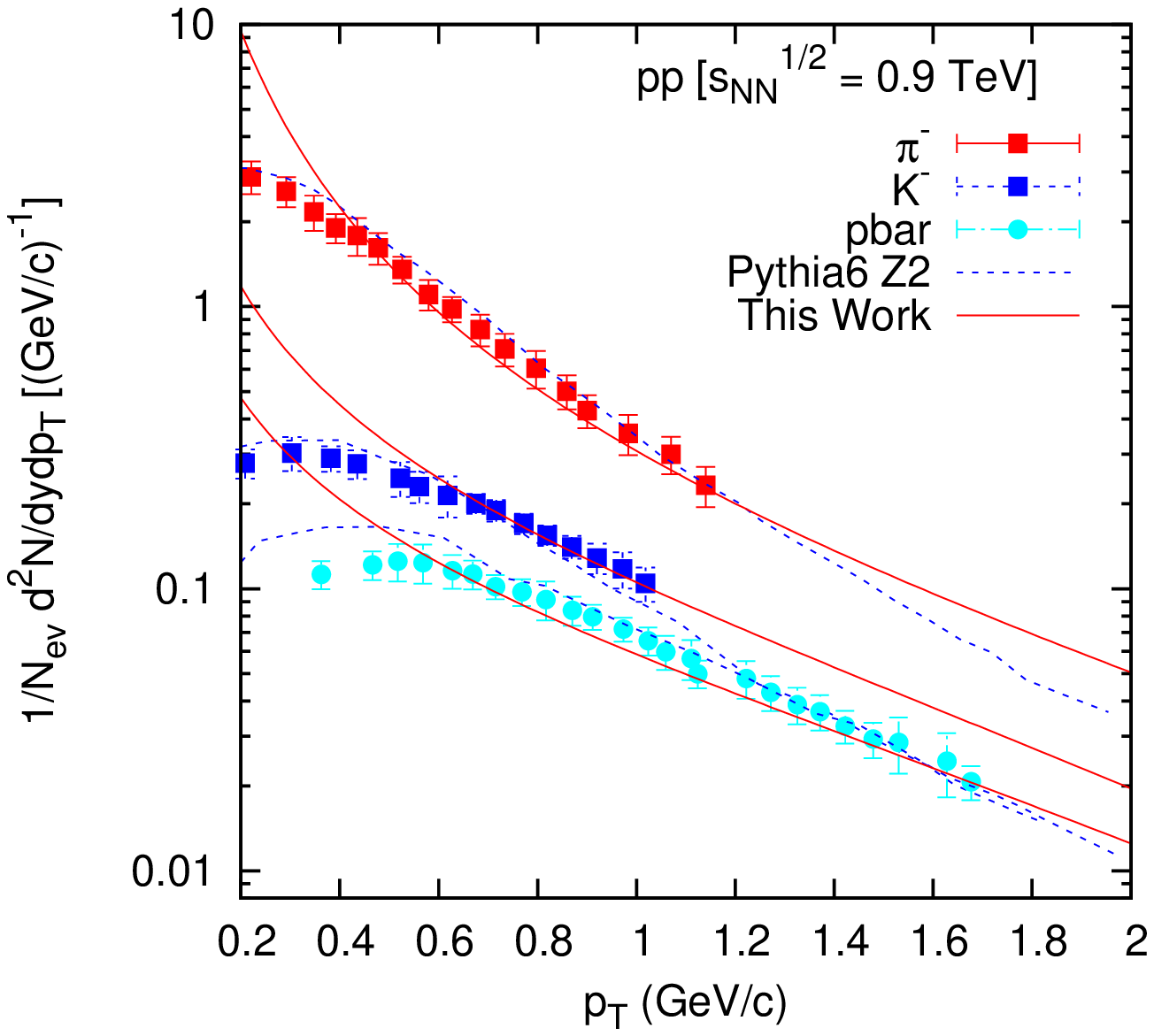}
\setcaptionwidth{2.6in}
\end{minipage}}%
\subfigure[]{
\begin{minipage}{0.5\textwidth}
\centering
 \includegraphics[width=2.5in]{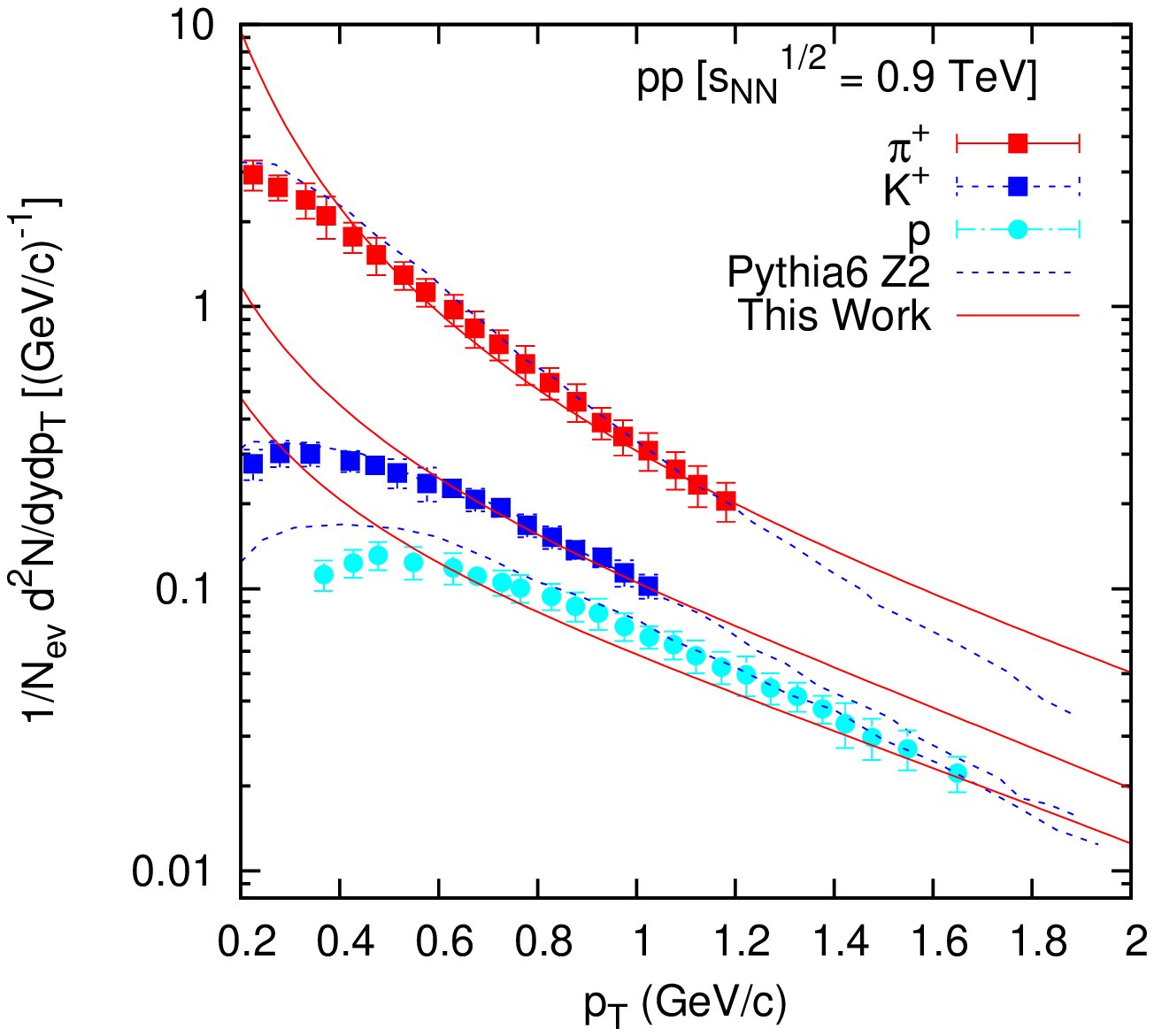}
  \end{minipage}}%
  \vspace{0.01in}
 \subfigure[]{
 \begin{minipage}{.5\textwidth}
\centering
 \includegraphics[width=2.5in]{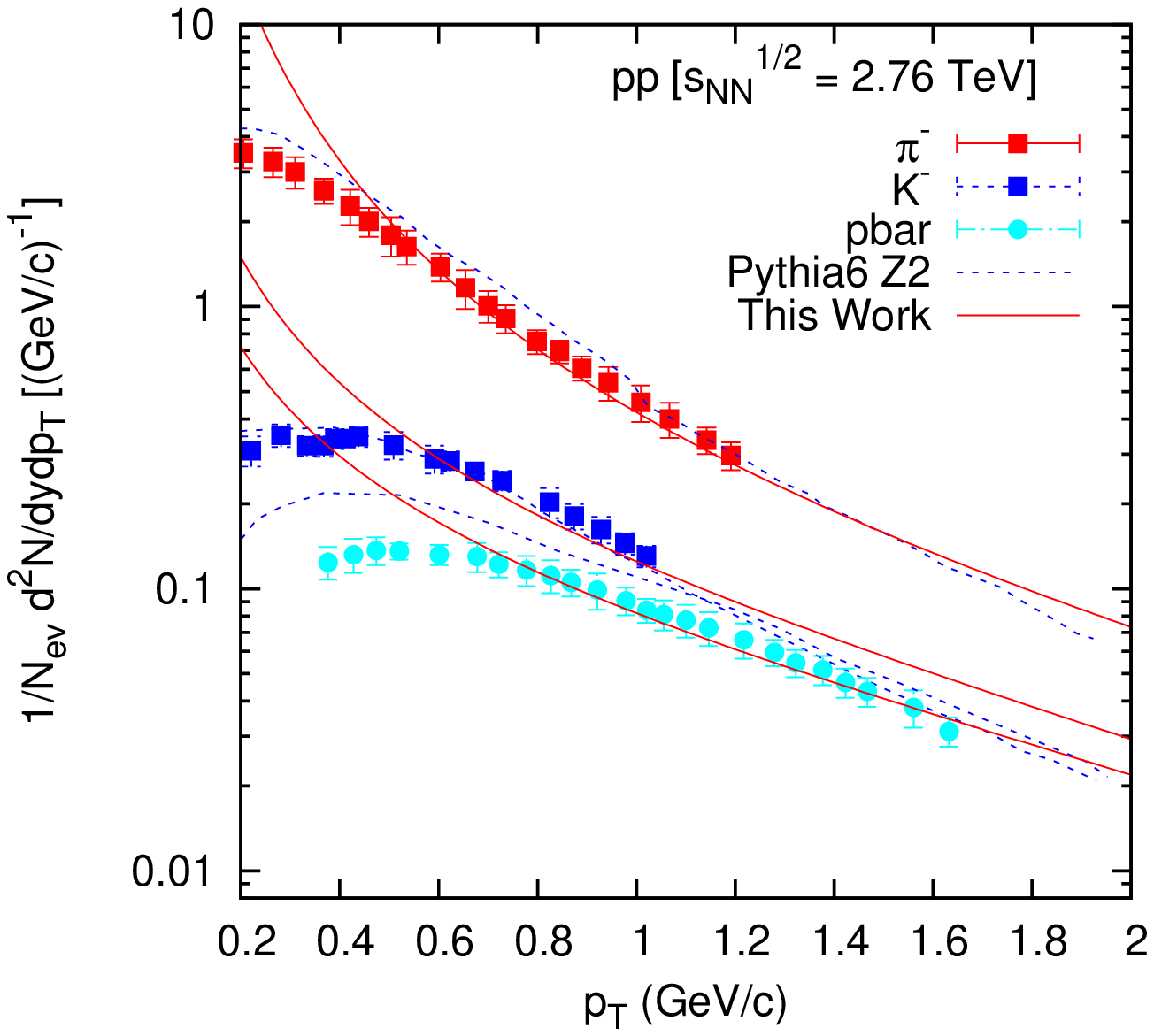}
 \setcaptionwidth{2.6in}
\end{minipage}}%
\subfigure[]{
\begin{minipage}{0.5\textwidth}
\centering
 \includegraphics[width=2.5in]{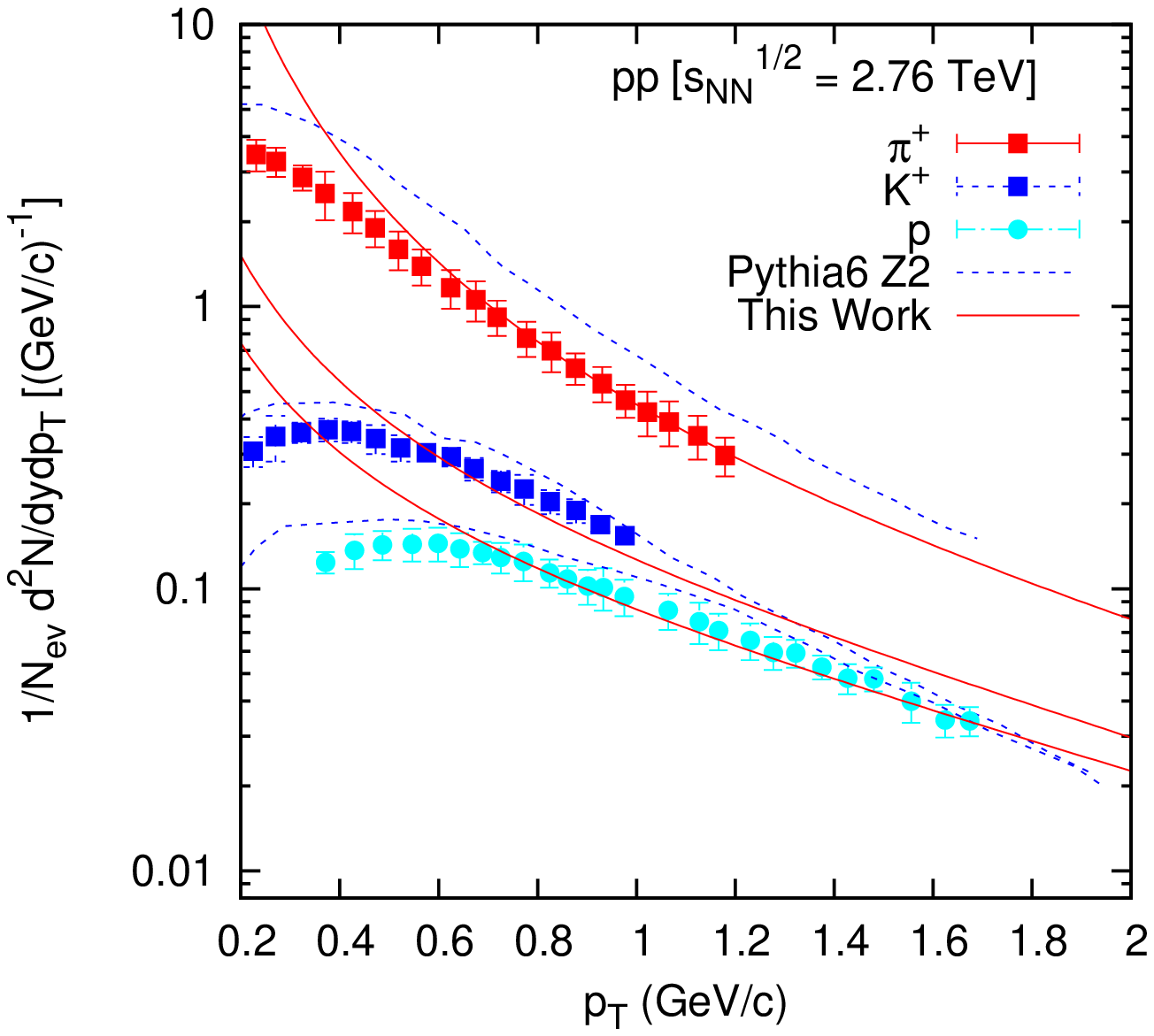}
  \end{minipage}}%
  \vspace{0.01in}
 \subfigure[]{
 \begin{minipage}{.5\textwidth}
\centering
 \includegraphics[width=2.5in]{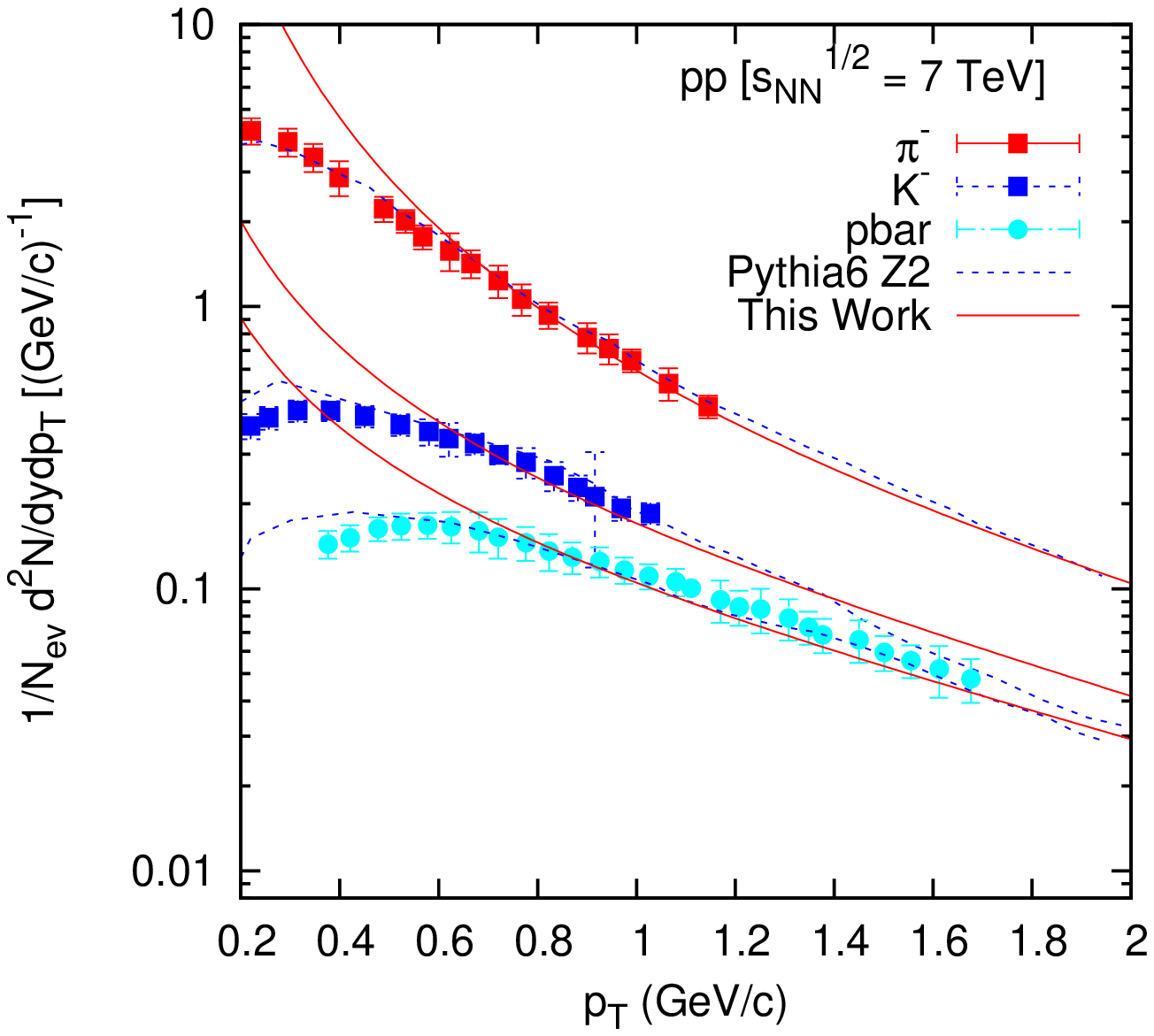}
 \setcaptionwidth{2.6in}
\end{minipage}}%
\subfigure[]{
\begin{minipage}{0.5\textwidth}
\centering
 \includegraphics[width=2.5in]{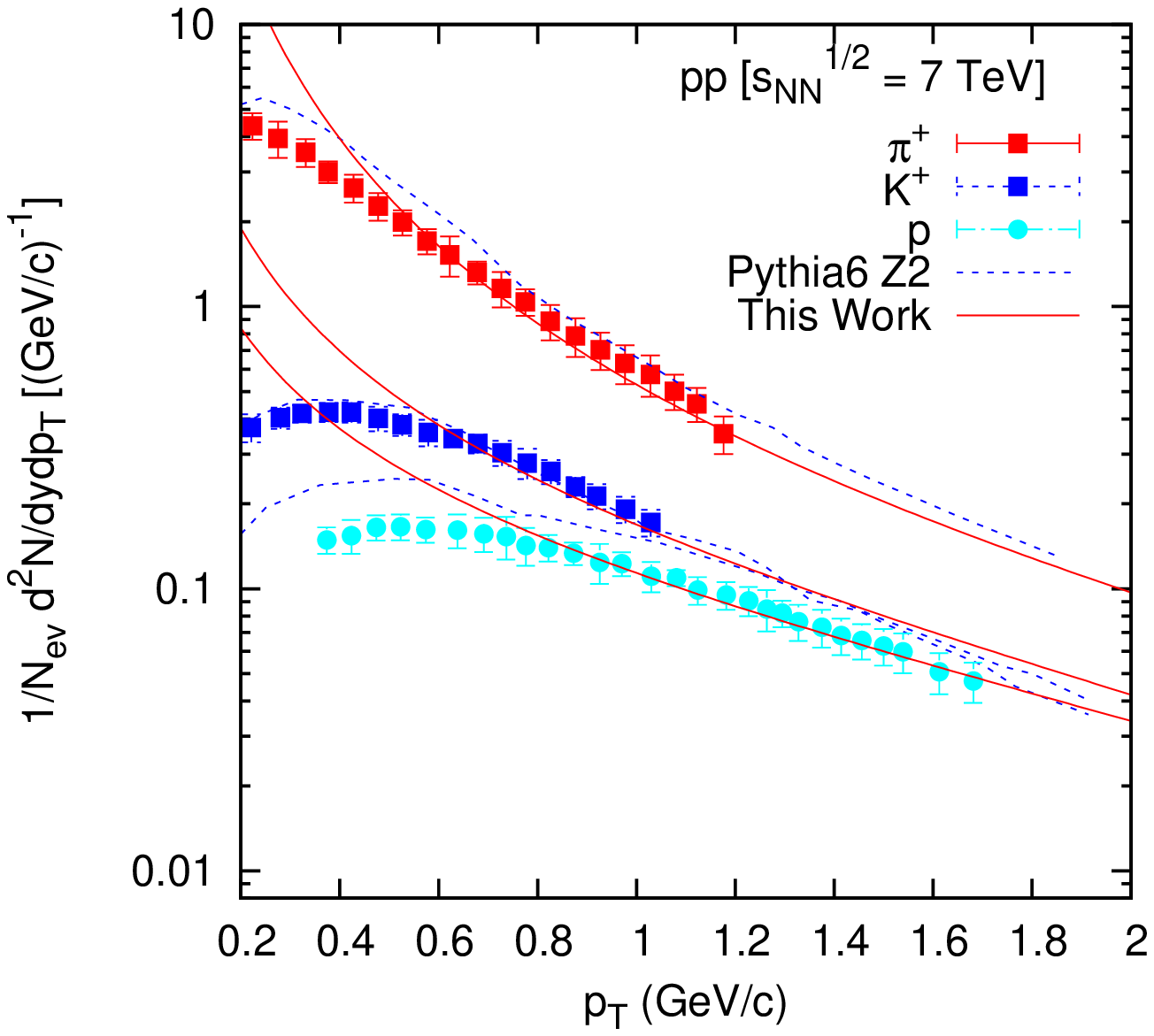}
  \end{minipage}}%
\caption{{\small Plots for $\pi$, $K$, $\bar p$ and $p$ productions in $p+p$ collisions at
energies $\sqrt{s_{NN}}$ = 0.9 TeV, $\sqrt{s_{NN}}$ = 2.76 TeV
and $\sqrt{s_{NN}}$ = 7 TeV. Data are taken \cite{cms12}. Solid lines in the Figures show the SCM-based
theoretical plots while the dotted ones show PYTHIA-based results \cite{cms12}.} }
\end{figure}

\newpage
\begin{figure}
\subfigure[]{
\begin{minipage}{.5\textwidth}
\centering
\includegraphics[width=2.5in]{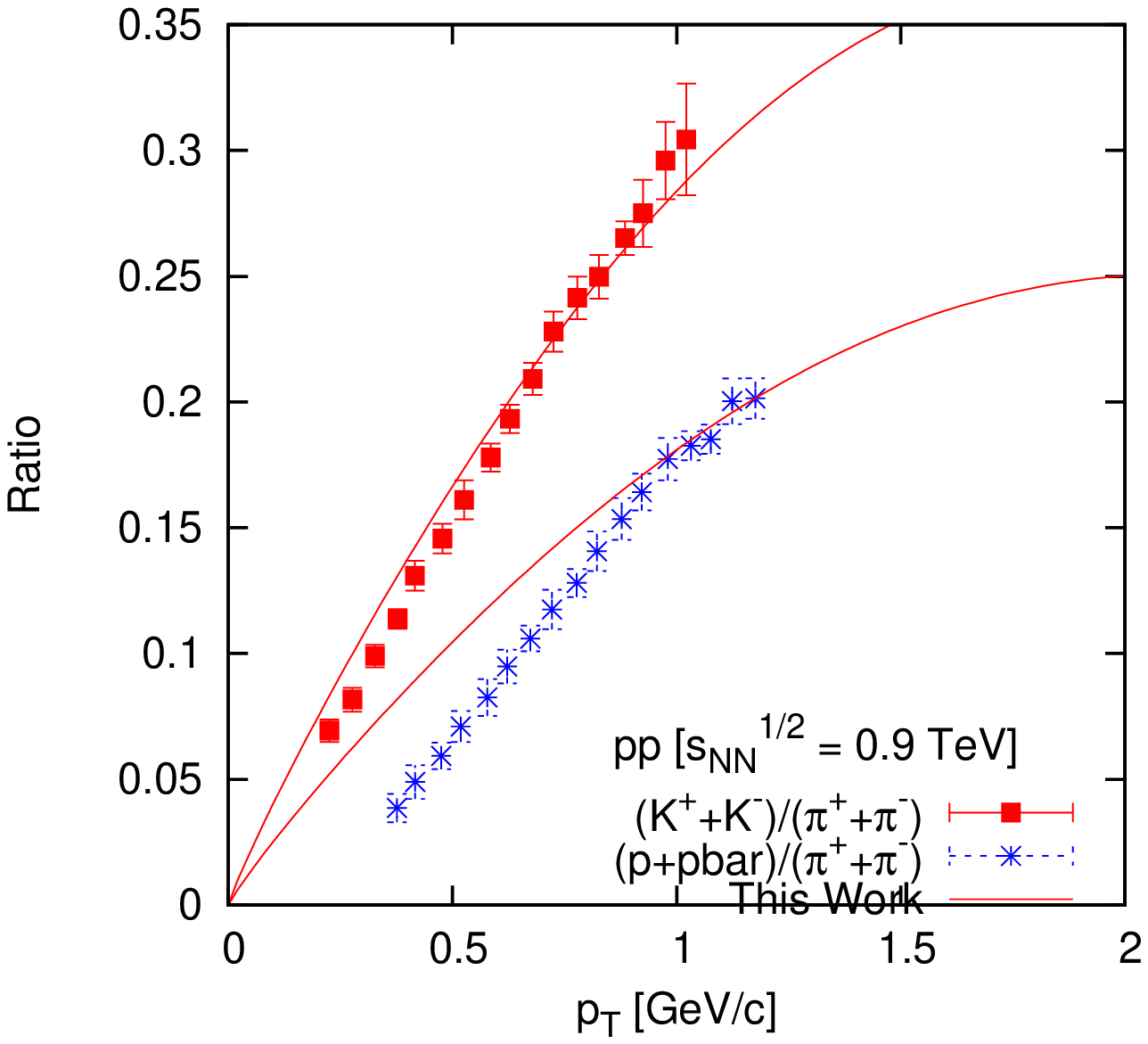}
\setcaptionwidth{2.6in}
\end{minipage}}%
\subfigure[]{
\begin{minipage}{0.5\textwidth}
\centering
 \includegraphics[width=2.5in]{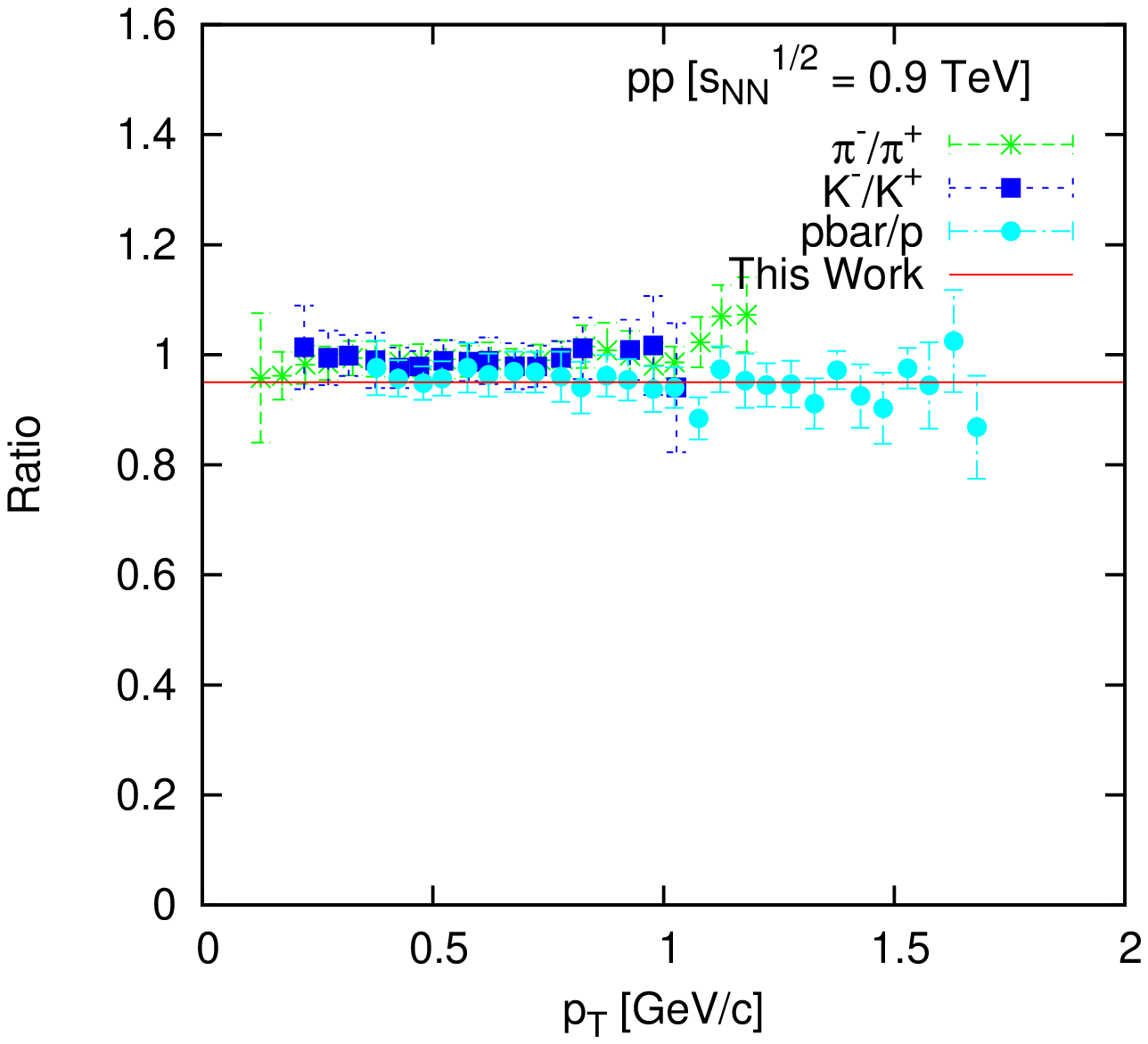}
  \end{minipage}}%
  \vspace{0.01in}
 \subfigure[]{
 \begin{minipage}{.5\textwidth}
\centering
 \includegraphics[width=2.5in]{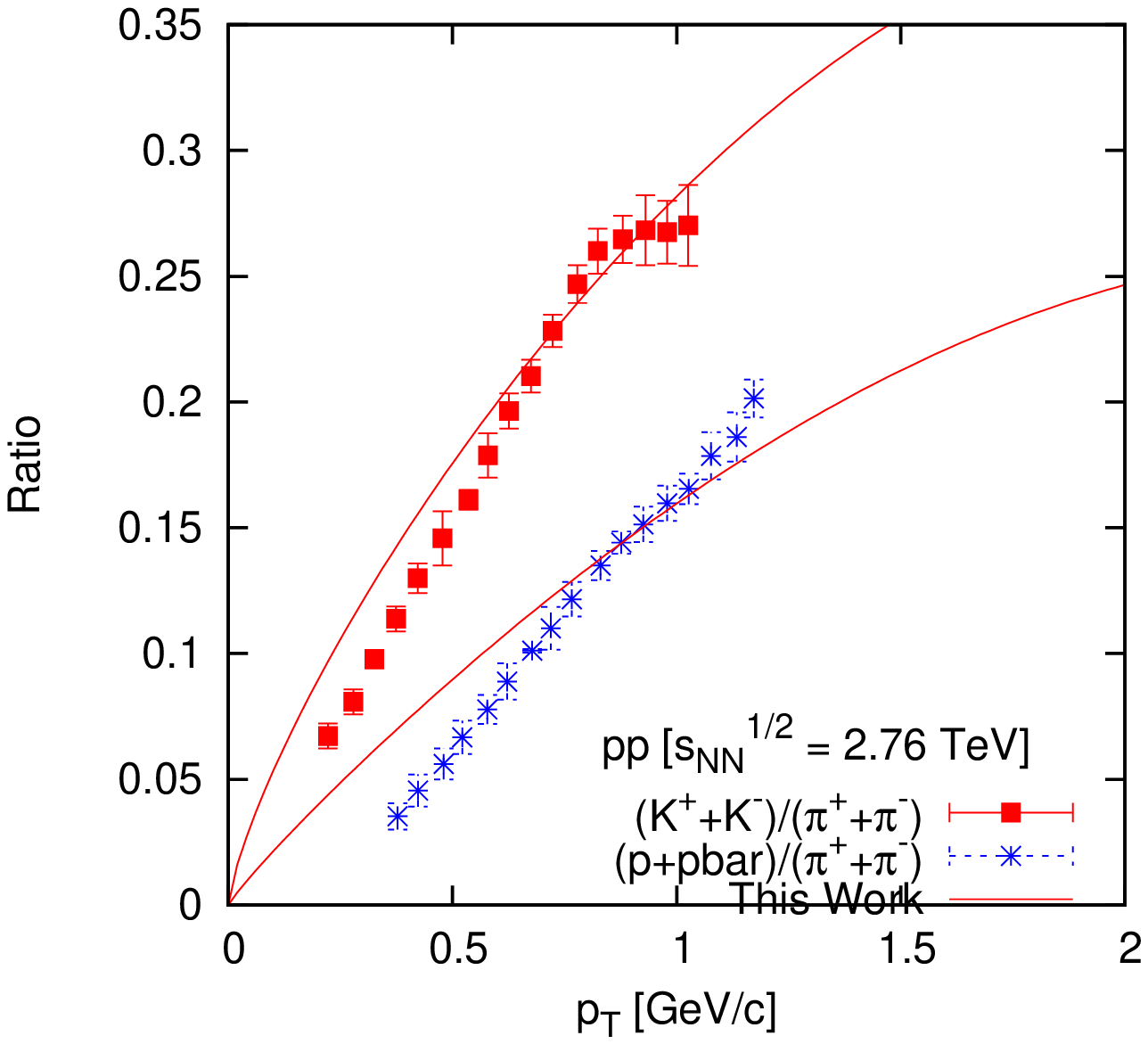}
 \setcaptionwidth{2.6in}
\end{minipage}}%
\subfigure[]{
\begin{minipage}{0.5\textwidth}
\centering
 \includegraphics[width=2.5in]{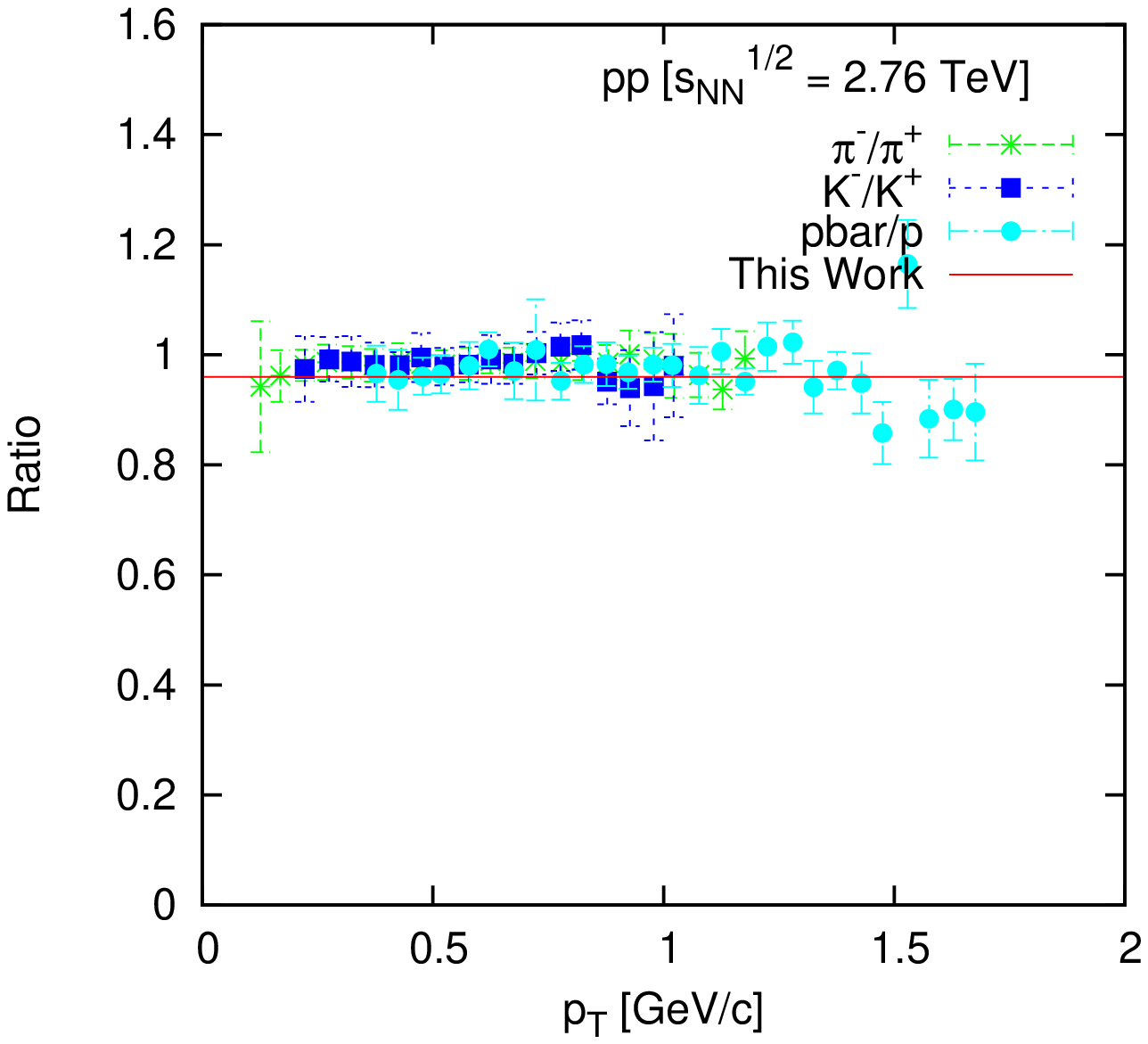}
  \end{minipage}}%
  \vspace{0.01in}
 \subfigure[]{
 \begin{minipage}{.5\textwidth}
\centering
 \includegraphics[width=2.5in]{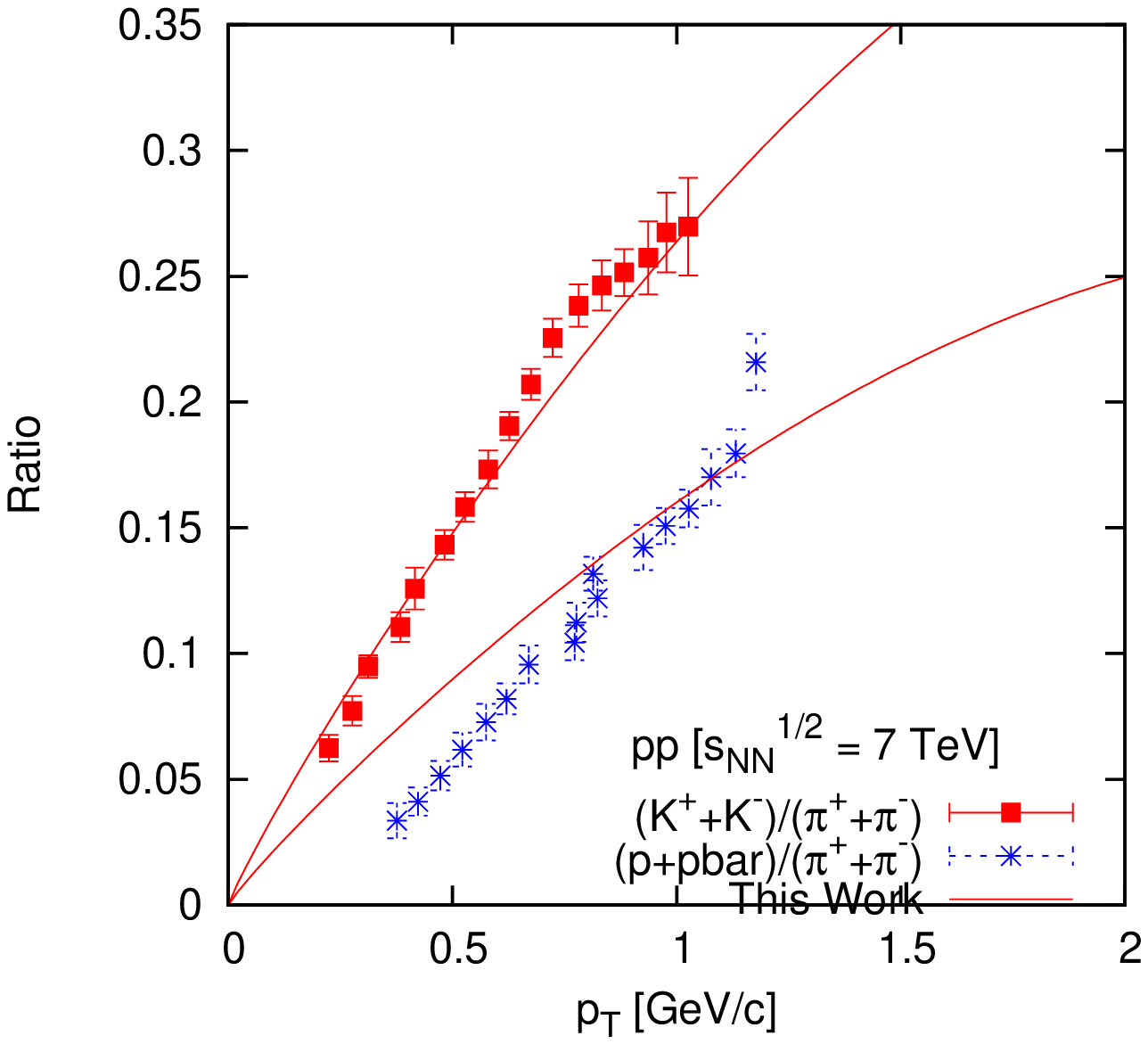}
 \setcaptionwidth{2.6in}
\end{minipage}}%
\subfigure[]{
\begin{minipage}{0.5\textwidth}
\centering
 \includegraphics[width=2.5in]{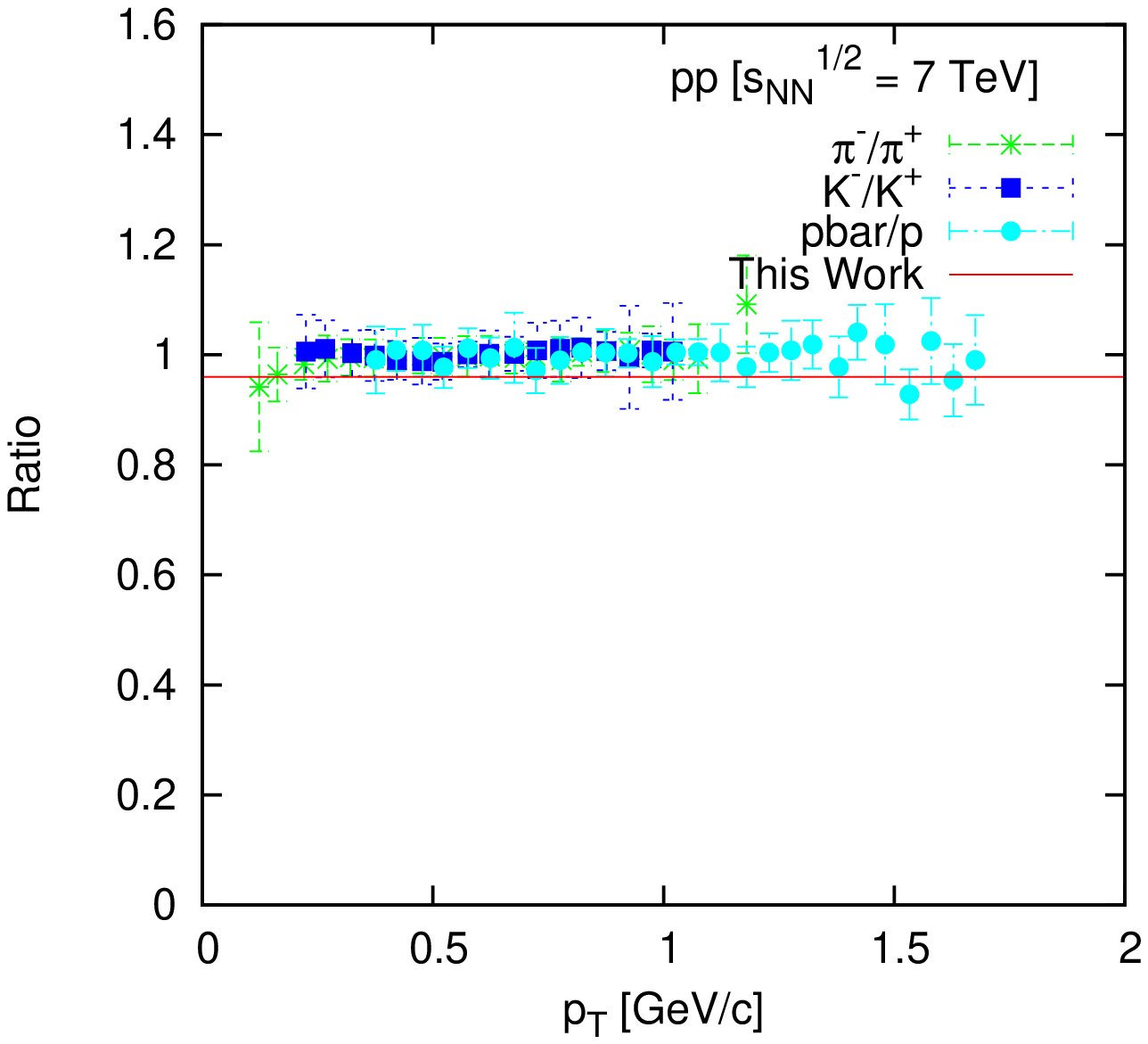}
  \end{minipage}}%
\caption{\small  Ratio behaviours for different hadrons at different LHC-energies. The figures in the left panel depict $K/\pi$ and $p/\pi$ ratios whereas the right panel shows $\pi^-/\pi^+$, $K^-/K^+$ and $p/{\bar p}$ ratio behaviour at
energies $\sqrt{s_{NN}}$ = 0.9 TeV, $\sqrt{s_{NN}}$ = 2.76 TeV
and $\sqrt{s_{NN}}$ = 7 TeV. Data are taken from \cite{cms12}, \cite{preg}. Lines in the Figures show the SCM-based
theoretical plots.}
\end{figure}

\end{document}